\documentclass[journal]{IEEEtran}
\usepackage{graphicx}
\usepackage{color}
\usepackage{placeins}
\usepackage{float}
\usepackage{tabularx,colortbl}
\usepackage{amssymb}
\usepackage{amsthm}
\usepackage{cite}
\usepackage{amsmath}

\makeatletter                               % algorithm
\newif\if@restonecol
\makeatother

\makeatletter
\newif\if@restonecol
\makeatother

\usepackage[linesnumbered,ruled,vlined]{algorithm2e}%[ruled,vlined]{
\usepackage{algpseudocode}
\usepackage{amsmath}
\usepackage{listings}                       % Matlab code
\usepackage[framed,numbered,autolinebreaks,useliterate]{mcode}
\theoremstyle{plain}

\theoremstyle{plain}

\usepackage{amssymb}
\usepackage{bbm}
\usepackage{amsmath,bm}

\usepackage{subfigure}
\usepackage{makecell}
\usepackage{threeparttable}
\usepackage{pdfsync}

\IEEEoverridecommandlockouts

\begin{document}
%%%%%%%%%%%%%%%%%%%%%%%%%%%%%%%%%%%%%%%%%%%%%%%%%%%%%%%%%%%%%%%%%%%%%%%
%----------------------------title&author&thanks----------------------------

\title{Channel Estimation for XL-MIMO Systems with Polar-Domain Multi-Scale Residual Dense Network}
\author{Hao Lei, Jiayi Zhang,~\IEEEmembership{Senior Member,~IEEE}, Huahua Xiao, Xiaodan Zhang,\\ Bo Ai,~\IEEEmembership{Fellow,~IEEE}, and  Derrick Wing Kwan Ng,~\IEEEmembership{Fellow,~IEEE}
\thanks{
H. Lei and J. Zhang are with the School of Electronic and Information Engineering, Beijing Jiaotong University, Beijing 100044, China  (e-mail: \{19211425, jiayizhang\}@bjtu.edu.cn).

H. Xiao is with ZTE Corporation and State Key Laboratory of Mobile Network and Mobile Multimedia Technology. (e-mail: xiao.huahua@zte.com.cn).

X. Zhang is with School of Management, Shenzhen Institute of Information Technology, Shenzhen 518172, China (e-mail: zhangxd@sziit.edu.cn).

B. Ai is with the State Key Laboratory of Rail Traffic Control and Safety, Beijing Jiaotong University, Beijing 100044, China (e-mail: boai@bjtu.edu.cn).

D. W. K. Ng is with the School of Electrical Engineering and Telecommunications, University of New South Wales, Sydney, NSW 2052, Australia (e-mail: w.k.ng@unsw.edu.au).
}
\vspace{-2em}}
\maketitle

%%%%%%%%%%%%%%%%%%%%%%%%%%%%%%%%%%%%%%%%%%%%%%%%%%%%%%%%%%%%%%%%%%%%%%%
%----------------------------abstract----------------------------
%%%%%%%%%%%%%%%%%%%%%%%%%%%%%%%%%%%%%%%%%%%%%%%%%%%%%%%%%%%%%%%%%%%%%%%
\begin{abstract}
Extremely large-scale multiple-input multiple-output (XL-MIMO) is a promising technique to enable versatile applications for future wireless communications. % , thanks to the utilization of extremely large number of antennas.
%To realize the huge potential performance gain, accurate channel state information is a fundamental technical prerequisite.
In conventional massive MIMO, the channel is often modeled by the far-field planar-wavefront with rich sparsity in the angular domain that facilitates the design of low-complexity channel estimation.
However, this sparsity is not conspicuous in XL-MIMO systems due to the non-negligible near-field spherical-wavefront.
To address the inherent performance loss of the angular-domain channel estimation schemes, we first propose the polar-domain multiple residual dense network (P-MRDN) for XL-MIMO systems based on the polar-domain sparsity of the near-field channel by improving the existing MRDN scheme.
Furthermore, a polar-domain multi-scale residual dense network (P-MSRDN) is designed to improve the channel estimation accuracy.
Finally, simulation results reveal the superior performance of the proposed schemes compared with existing benchmark schemes and the minimal influence of the channel sparsity on the proposed schemes.
\end{abstract}

%%%%%%%%%%%%%%%%%%%%%%%%%%%%%%%%%%%%%%%%%%%%%%%%%%%%%%%%%%%%%%%%%%%%%%%
%----------------------------keywords----------------------------
%%%%%%%%%%%%%%%%%%%%%%%%%%%%%%%%%%%%%%%%%%%%%%%%%%%%%%%%%%%%%%%%%%%%%%%


%\newpage
\IEEEpeerreviewmaketitle
\vspace{-0.3cm}
%%%%%%%%%%%%%%%%%%%%%%%%%%%%%%%%%%%%%%%%%%%%%%%%%%%%%%%%%%%%%%%%%%%%%%%
%----------------------------introduction----------------------------
%%%%%%%%%%%%%%%%%%%%%%%%%%%%%%%%%%%%%%%%%%%%%%%%%%%%%%%%%%%%%%%%%%%%%%%
\section{Introduction}
To significantly improve the required ultra-low access latency and ultra-high data rate of the sixth-generation (6G) wireless communications, extremely large-scale multiple-input multiple-output (XL-MIMO) has been considered as one of the promising techniques \cite{[1]}, \cite{[2]}.
The general idea of XL-MIMO is to deploy another order-of-magnitude antenna number (e.g., $512$ or more) at the base station (BS), compared with only $64$ or $128$ antennas in those conventional massive MIMO (mMIMO) systems \cite{[1]}, \cite{[2]}.
In this way, XL-MIMO can provide higher spatial degrees-of-freedom (DoF) \cite{[1]} and spectral efficiency (SE) \cite{[14]} compared with mMIMO systems.

To achieve the desired performance in XL-MIMO networks in practice, several challenges must be addressed, e.g., accurate channel modeling, low-complexity signal processing, spatial non-stationary characteristics, etc.
Among these challenges, channel estimation (CE) is a critical one as accurate channel state information (CSI) is a fundamental requirement for effective signal processing.
First of all, the channel between the BS and its user equipment (UE) is with an extremely high dimensionality due to the deployment of large-scale antennas.
Secondly, the operating region of XL-MIMO shifts from far-field to near-field \cite{[16]}, where the boundary between near-field and far-field is defined by the Rayleigh distance.
With a high carrier frequency, the Rayleigh distance can be extended to hectometre-range or even longer, \cite{[16]}, \cite{[15]}, which means that XL-MIMO channels should be modeled by near-field spherical waves.
The difference in electromagnetic (EM) characteristics results in different sparsity properties.
Thus, the direct application of angular-domain CE schemes will inevitably suffer from a degradation in normalized mean square error (NMSE) performance for XL-MIMO \cite{[3]}, \cite{[5]}.
%Consequently, XL-MIMO channels no longer exhibit rich sparsity in the angular domain \cite{[3]}.
%Last but not least, spatial non-stationary characteristics also will make CE a significant challenge \cite{[1]}.

Recently, several works have focused on CE problems in XL-MIMO systems \cite{[3]}-[11].
For instance, the authors in \cite{[3]} proposed a polar-domain representation for the XL-MIMO channel that fully captures the near-field spherical wave characteristics.
It is noteworthy that compressed sensing (CS) algorithms can be exploited to perform CE in XL-MIMO systems with acceptable NMSE performance based on the polar-domain channel sparsity \cite{[3]}-\!\!\cite{[6]}.
Additionally, CS-based CE schemes have shown reasonable performance for XL-MIMO networks with spatial non-stationary properties \cite{[19]}, \cite{[20]}.
%The least-squares (LS) scheme, which requires no prior information and has low computational complexity, was adopted for CE in the XL-MIMO.
Moreover, a reduced-subspace least-squares (RS-LS) CE scheme was proposed based on the least-squares (LS) estimator by utilizing the compact eigenvalue decomposition of the spatial correlation matrix  \cite{[7]}.
Despite various efforts have been devoted to CE, satisfactory estimation accuracy is yet to be achieved due to the limited available resources.

In recent years, deep learning (DL)-based CE algorithms have gained significant momentum.
For instance, in reconfigurable intelligent surface (RIS)-aided mMIMO systems, a multiple residual dense network (MRDN) was designed for CE with high estimation accuracy by exploiting the angular-domain channel sparsity \cite{[8]}.
Also, the authors in \cite{[25]} proposed a U-shaped multilayer perceptron (U-MLP) network to estimate the near-field channel with spatial non-stationary properties by capturing the long-range dependency of channel features.
In addition, deep learning networks have been exploited to extract the parameters of near-field channels, which can be adopted to reconstruct the channel matrices \cite{[9]}, \cite{[23]}.
Moreover, the authors in \cite{[24]} formulated a near-field CE problem as a compressed sensing problem and then proposed a sparsifying dictionary learning-learning iterative shrinkage and thresholding algorithm (SDL-LISTA) by formulating the sparsifying dictionary as a neural network layer.
Despite these advances, there is still a lack of sufficient investigations to reveal the impact of inherent near-field channel sparsity in different domains on DL-based CE schemes.

%However, directly applying the MRDN-based CE scheme to XL-MIMO is expected to suffer from certain performance loss due to different sparsity properties.
In this paper, we propose a polar-domain multiple residual dense network (P-MRDN)-based CE scheme to explicitly exploit the polar-domain channel sparsity in XL-MIMO systems and evaluate the NMSE performance of the MRDN and P-MRDN-based CE schemes.
However, the existing DL-based CE schemes for near-field did not consider the multi-scale feature, which generally limits their accuracy.
To address this issue, inspired by \cite{[13]}, the notion of atrous spatial pyramid pooling (ASPP), which adopts parallel atrous convolution layers with different rates to capture the multi-scale information \cite{[12]}, is incorporated into the proposed P-MRDN to further improve the CE accuracy.
It is worth noting that our proposed schemes are expected to outperform the state-of-the-art CE schemes in NMSE due to the tailor-mode approach.
The main contributions can be summarized as follows.
\begin{itemize}
  \item We propose a P-MRDN-based CE scheme for XL-MIMO systems by exploiting the polar-domain channel sparsity. More importantly, we reveal
      the impact of the channel sparsity in different domains on DL-based CE schemes. %the rich sparsity of XL-MIMO channels in the polar domain.
  \item Atrous spatial pyramid pooling-based residual dense network (ASPP-RDN) is also proposed by exploiting ASPP as a parallel branch of RDN. Then, a  polar-domain multi-scale residual dense network (P-MSRDN)-based CE scheme is proposed to further improve the estimation accuracy based on ASPP-RDN. %Finally, numerical results demonstrate the effectiveness of the proposed schemes.
  \item Numerical results demonstrate that the performance of the proposed schemes can significantly outperform existing state-of-the-art CE schemes\footnote{ Simulation codes are provided to reproduce the results in this paper: https://github.com/BJTU-MIMO.}.%conventional schemes and existing polar-domain
      %Numerical results reveal the superiority of the proposed schemes in NMSE performance compared with the state-of-the-art CE schemes.
%  \item We derive exact achievable uplink SE expressions for two signal processing schemes with arbitrary combining scheme.\footnote{ Simulation codes are provided to reproduce the results in this paper: https://github.com/BJTU-MIMO.}
\end{itemize}

{\textbf{\textit{Notation}}:}
Boldface lowercase letters $\rm {\bf{a}}$ and boldface uppercase letters $\rm {\bf{A}}$ denote column vectors and matrices, respectively.
Transpose is denoted by $(\cdot)^{T}$.
%Conjugate, transpose, and conjugate transpose are denoted by $(\cdot)^{*}$, $(\cdot)^{T}$ and $(\cdot)^{H}$.
We denote the $ M \times N $ complex-valued matrix and $ M \times N $ real-valued matrix by $\mathbb{C}^{ M \times N}$ and $\mathbb{R}^{ M \times N}$, respectively.
We adopt $\mathbb{E} \{ \cdot \} $  to denote the expectation operator.
The circularly symmetric complex Gaussian distribution with covariance $ \sigma^2 $ and the uniform distribution between $a$ and $b$ are denoted by $\mathcal{CN}(0,\sigma^2)$ and $\mathcal{U} (a,b)$, respectively.
The Euclidean norm is denoted by $ \|\cdot\| $.
%$ \mathcal{O}(\cdot) $ denotes the computational complexity.
%The definitions and the determinant of a matrix are denoted by $ \triangleq$ and $ \left| \cdot \right| $, respectively.
%$\otimes$ and $\odot$ denote the Kronecker products and the element-wise products, respectively.
%${\rm \bf{I}}_n $ is the $n\times n$ identity matrix, and $ {\rm \bf{1}}_n$ is a column vector with all ones.
%The circularly symmetric complex Gaussian distribution is denoted by $\mathcal{CN}(0,\sigma^2)$.

\begin{figure}
  \centering
    \setlength{\abovecaptionskip}{-0cm}
  \includegraphics[width=2.7in]{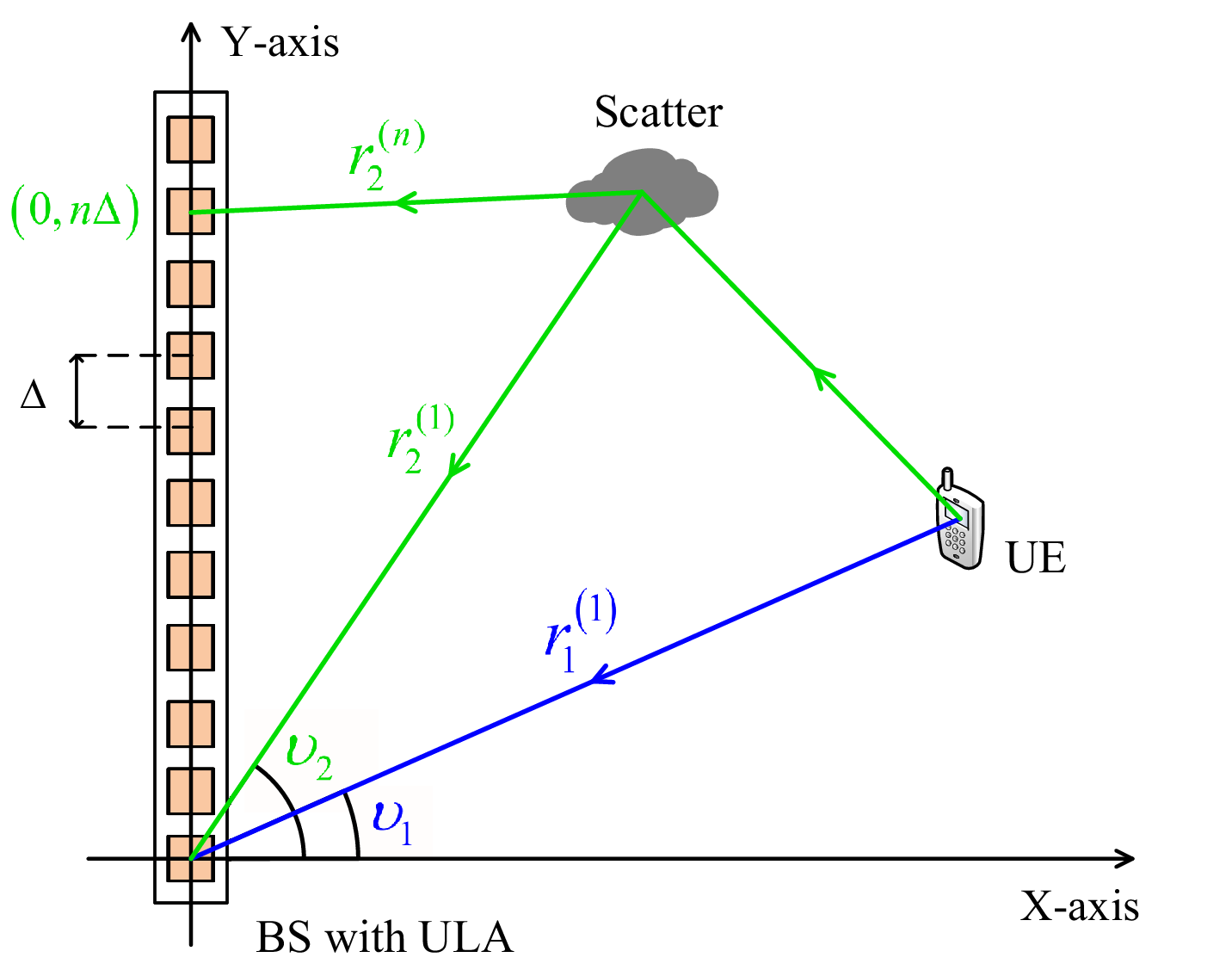}
        %\vspace{-0.2em}
  \caption{Illustration of the XL-MIMO system with a uniform linear array (ULA)-based BS, where the UE and scatter are located in near-field. The figure depicts the angle $ v_l $ and the distance $ r_l^{(n)}$ between the $n$-th BS antenna, $ \forall n \in \{1,\cdots,N\} $, and the scatter or the user for the $l$-th path, where $ \Delta $ represents the antenna spacing. }\label{ULA-XL-MIMO}
  \vspace{-0.4cm}
\end{figure}

%%%%%%%%%%%%%%%%%%%%%%%%%%%%%%%%%%%%%%%%%%%%%%%%%%%%%%%%%%%%%%%%%%%%%%%
%-----------------------system and channel modeling--------------------
%%%%%%%%%%%%%%%%%%%%%%%%%%%%%%%%%%%%%%%%%%%%%%%%%%%%%%%%%%%%%%%%%%%%%%
 %\vspace{-0.3cm}
\section{System Model}

%By supposing the coordinate of the $n$-th antenna is $ (0, {{\delta _n}\Delta} )$, where $ \delta _n = \frac{2n-N+1}{2}, n = 0,1,\cdots,N-1 $,

We consider an uplink time division duplexing (TDD) XL-MIMO system.
As illustrated in Fig. 1, we consider a uniform linear array (ULA)-based BS with $N$ antennas and one single-antenna UE for the XL-MIMO system\footnote{As for multi-UE scenarios, the channels between the BS and its different UEs can be modeled separately by the spherical-wave assumption.}.
The antenna spacing is denoted by $ \Delta = {\lambda  \mathord{\left/ {\vphantom {\lambda  2}} \right. \kern-\nulldelimiterspace} 2} $, where $ \lambda $ is the carrier wavelength.
We denote the channel between the UE and the BS by $ {\rm {\bf h}} \in \mathbb{C}^{ N \times 1}$, where $ N $ is the number of antennas at the BS.
By assuming that the UE sends a predefined pilot sequence, set as 1 for simplicity, we can represent the received signal $ {\rm{\bf y}} \in { {\mathbb{C}}^{ N  \times 1}} $ at the BS as
\begin{equation}\label{received signal}
  {\rm{\bf y}} = \sqrt p {\rm{\bf h}} + {\rm{\bf n}},
\end{equation}
where $p$ is the transmit power of the UE, $ {\rm{\bf n}}  \in {\mathbb{C}^{N  \times 1}} $ denotes the receiver noise with independent $ \mathcal{C}\mathcal{N}(0,\sigma^2)  $ entries, and $ \sigma^2 $ denotes the noise power.
To further unveil the channel sparsity in XL-MIMO, the channel modelings for far-field and near-field are reviewed as follows:

$\bullet$ \emph{Far-field channel modeling}:
In conventional far-field region, the channel is modeled by the planar wave, which can be expressed as
%which is the superposition of one line-of-sight (LoS) path and $L-1$ non-line-of-sight (NLoS) paths, i.e.,
\begin{equation}\label{far-field channel}
  {\rm  {\bf h} }^{\rm far-field} = \sqrt{ \frac{N}{L} } \sum\limits_{l = 1}^L {{\beta_l}{e^{ - jk{r_l}}} {\rm{\bf a}}( {{\theta _l}})},
\end{equation}
where $ k = \frac{2 \pi}{\lambda} $ is the wave number.
We assume that there is one line-of-sight (LoS) path and $L-1$ non-line-of-sight (NLoS) paths \cite{[3]}-[8].
%$L$ denotes the number of paths,
Moreover, we denote the angle, the distance, and the complex path gain of the $l$-th path by $ {\theta _l}= \sin \upsilon_l $, $ r_l $, and $ \beta_l $, respectively.
%Moreover, the complex path gain, the distance, and the angle of $l$-th path are denoted by $ \beta_l $, $ r_l $ and $ {\theta _l}= \sin \upsilon_l $, respectively.
The steering vector $ {\rm{\bf a}}( {\theta _l} ) $ can be represented as
\begin{equation}\label{a}
  {\rm{\bf a}}( {\theta _l} ) = \frac{1}{{\sqrt N }}{\left[1, {   {e^{ j\pi{\theta _l} }}, \cdots , {e^{ j(N-1)\pi{\theta _l} }}   } \right]^T}.
\end{equation}
More interestingly, to exploit the angular-domain channel sparsity, the corresponding angular-domain representation $ {\rm  {\bf h} }^{\rm far-field}_{ \mathrm{A} }  $ can be derived from the channel $ {\rm  {\bf h} }^{\rm far-field} $ as \cite{[3]}
\begin{equation}\label{F}
  {\rm  {\bf h} }^{\rm far-field} = {\rm{\bf{F}}} {\rm  {\bf h} }^{\rm far-field}_{\mathrm{A}},
\end{equation}
where $ {\rm{\bf{F}}} = [ {\rm{\bf a}}( {{\theta _0}}), \cdots , {\rm{\bf a}}( {{\theta _{N-1}}}) ]  \in \mathbb{C}^{N \times N} $ is the Fourier transform matrix with $ \theta_n = \frac{2n - N + 1 }{N} , n = 0, 1, \cdots , N-1$.
Based on the angular-domain sparsity, several CS-based CE schemes have been proposed for far-field applications, e.g., \cite{[18]},  \cite{[21]}.

However, the angular-domain sparsity is not remarkable in XL-MIMO.
%However, the angular-domain channel sparsity is invalid in the XL-MIMO.
The reason is that the Rayleigh distance, $Z = {{2{D^2}} \mathord{\left/ {\vphantom {{2{D^2}} \lambda }} \right. \kern-\nulldelimiterspace} \lambda }$, can be in the range of hectometre or longer in XL-MIMO, where $D$ is the array aperture.
For instance, the Rayleigh distance is around $67$ meters with the array aperture of $1$ meters and the carrier frequency of $10$ GHz.
Therefore, we should consider the scenario that the UE is in near-field for XL-MIMO.

$\bullet$ \emph{Near-field channel modeling}:
Based on the exact spherical wave, the near-field channel can be expressed as \cite{[3]}
%Then, the XL-MIMO channel should be modeled by the exact spherical wave, which
\begin{equation}\label{near-field channel}
  {\rm{\bf h  }}^{\rm near-field} = \sqrt {\frac{N}{L}} \sum\limits_{l = 1}^L {{\beta_l}{e^{ - jk{r_l}}}{\rm{\bf b}}( {{\theta _l},{r_l}} )}.
\end{equation}
It is worth noting that the near-field steering vector $ {\rm{\bf b}}( {{\theta _l},{r_l}} ) $  is expressed as
\begin{equation}\label{b}
 \! {\rm{\bf b}}( {{\theta _l},{r_l}} ) \! =\!  \frac{1}{{\sqrt N }}  {\left[  1,  {{e^{ - j{k}\left( { r_l^{(2)} - {r_l^{(1)}}} \right)}}, \cdots ,{e^{ - j{k}\left( {r_l^{(N)} - {r_l^{(1)}}} \right)}}} \right]^T}.
\end{equation}
%where $ k = \frac{2 \pi}{\lambda} $ denotes the wavenumber.
We denote the distance between the $n$-th BS antenna and the UE or scatter by $ r_l^{(n)} $, as shown in Fig. 1.
Without loss of generality, we set the coordinate of the $n$-th antenna as $ (0, {  n\Delta} )$ such that $ r_l^{(n)} = \sqrt {{{\left( {{r_l^{(1)}}\sqrt {1 - \theta _l^2}  - 0} \right)}^2} + {{\left( {{r_l^{(1)}}{\theta _l} - n\Delta} \right)}^2}}  = \sqrt { \left({r_l^{(1)}}\right)^2 + n^2{\Delta^2} - 2{r_l^{(1)}}{\theta _l}{n}\Delta}  $, where $ \theta _l=\sin \upsilon_l \in [-1,1] $ denotes the spatial angle.
%By supposing the coordinate of the $n$-th antenna is $ (0, {{\delta _n}\Delta} )$ with $ \delta _n = \frac{2n-N+1}{2}, n = 0,1,\cdots,N-1 $,, we can drive that $ r_l^{(n)} = \sqrt {{{\left( {{r_l^{(1)}}\sqrt {1 - \theta _l^2}  - 0} \right)}^2} + {{\left( {{r_l^{(1)}}{\theta _l} - {\delta _n}\Delta} \right)}^2}}  = \sqrt { ({r_l^{(1)}})^2 + \delta _n^2{\Delta^2} - 2{r_l^{(1)}}{\theta _l}{\delta _n}\Delta}  $, where $ \theta _l=\sin \upsilon_l \in [-1,1] $ denotes the spatial angle.

Note that the angular field distribution is independent of the distance under the planar-wave assumption in the far-field, as shown in \eqref{a}.
By contrast, the EM waves in XL-MIMO systems should be modeled by the spherical-wave assumption, showing distance-dependent angular field distribution, as shown in \eqref{b}.
%(i.e., the polar field distribution), as shown in \eqref{b}.
%the XL-MIMO operating in the near-field under the spherical-wave assumption displays distance-dependent angular field distribution, i.e., the polar field distribution.
%More specifically, the far-field steering vector $ {\rm{\bf a}}(\cdot) $ only depends on the angel $ \theta _l $, while the near-field steering vector $ {\rm{\bf b}}(\cdot) $ depends on both the angel $ \theta _l $ and the distance $ r_l^{(n)} $.

To fully leverage the near-field channel sparsity, a polar-domain transform matrix $ \rm { \bf D} $ was proposed in \cite{[3]}, which is denoted by $ {\rm{\bf D}} =  [  {\rm{\bf D}}_{Q_1},{\rm{\bf D}}_{Q_2},\cdots, {\rm{\bf D}}_{Q_N}   ]    $  with $   {\rm{\bf D}}_{Q_n}    =  [ {{\rm{\bf b}}\left( {{\theta _n},{r_{n}^{1}}} \right), \cdots ,{\rm{\bf b}}\left( {{\theta _n},{r_{n}^{Q_n}}} \right)} ]    $, where $ Q_n $ denotes the number of sampled distances at the sampled angle $ \theta _n $.
%Each column of $ {\rm{\bf D}} $ is a near-field steering vector with the sampled angle $ \theta _n $ and distance $ {r_{n}^{S_n}} $, where $ s_n = 1 , 2, \cdots, S_n $, $ S_n $ denotes the number of sampled distances at the sampled angle $ \theta _n $ \cite{[3]}.
Besides, we denote the number of all sampled distances by $ Q = \sum_{n=1}^{N}Q_n $.
%Besides, the number of all sampled distances can be denoted by $ Q = \sum_{n=1}^{N}S_n $.
Based on the matrix $ \rm { \bf D}  \in \mathbb{C}^{ N \times Q} $, the near-field channel is given by \cite{[3]}
\begin{equation}\label{polar}
   {\rm  {\bf h} }^{\rm near-field} = {\rm{\bf{D}}} {\rm  {\bf h} }^{\rm near-field}_{\mathrm{ P}}.
\end{equation}
%where  $ Q $ is the number of sampled near-field steering vectors \cite{[3]}.
Similar to the angular-domain sparsity of far-field channels, near-field channels exhibit certain sparsity in the polar domain.
% by the matrix $ {\rm{\bf{D}}}  $ which accounts for both the angle and distance information.
%Besides, since the matrix $ {\rm{\bf{D}}}  $ take into account both the angles and distances, the number of sampled near-field steering vectors $Q$ is generally larger than $N$.
%Besides, the number of sampled near-field steering vectors $Q$ is generally larger than $N$ due to the consideration of both the angles and distances in the matrix $ {\rm{\bf{D}}}  $.
Thus, CE can be performed with acceptable NMSE performance based on the polar-domain channel sparsity \cite{[3]}-[8].

\begin{figure}
	\addtocounter{figure}{0} %%%%%%%%%note1
	\centering
	\subfigure[MRDN-based channel estimation scheme.]{
		\begin{minipage}[t]{1\linewidth}%%%%%%%%%note2
        \centering
			\includegraphics[width=0.9\linewidth]{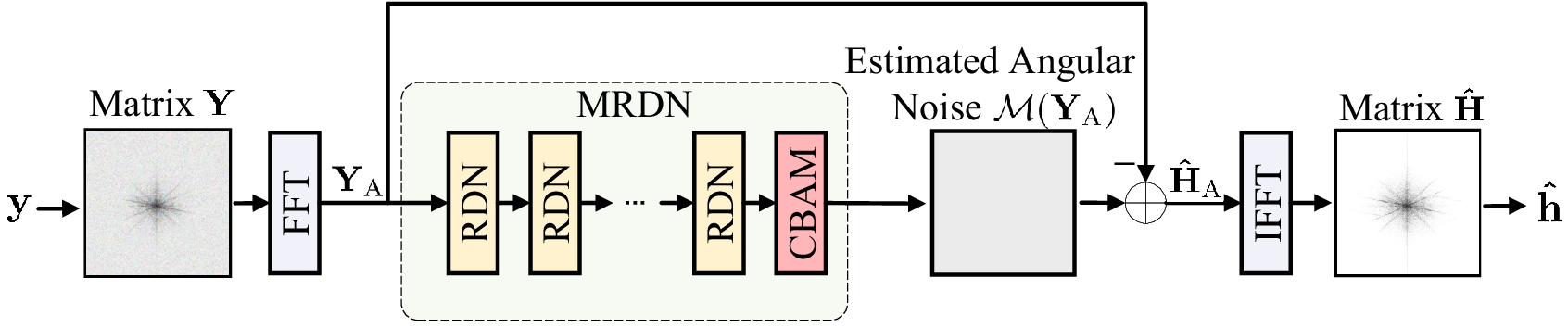}%%%%%%%%%note3
		
		\label{0}
		\end{minipage}%
	}%

%%%%%%%%%note4
	\subfigure[P-MRDN-based channel estimation scheme.]{
        \centering
		\begin{minipage}[t]{1\linewidth}
        \centering
				\includegraphics[width=0.9\linewidth]{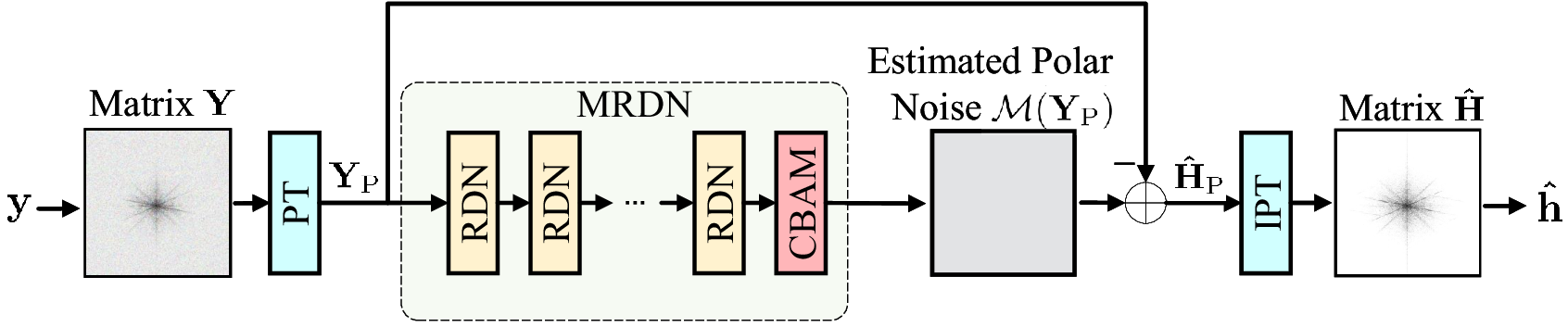}
			
			\label{1}
		\end{minipage}%
	}%
     % \vspace{-0.4em}
	\caption{Comparison between (a) MRDN-based channel estimation scheme and (b) PMRDN-based channel estimation scheme. The significant difference is the FFT in MRDN and the polar-domain transform (PT) in P-MRDN.}
	\centering
  \vspace{-0.3cm}
\end{figure}

%%%%%%%%%%%%%%%%%%%%%%%%%%%%%%%%%%%%%%%%%%%%%%%%%%%%%%%%%%%%%%%%%%%%%%%
%-----------------------Proposed channel estimation schemes--------------------
%%%%%%%%%%%%%%%%%%%%%%%%%%%%%%%%%%%%%%%%%%%%%%%%%%%%%%%%%%%%%%%%%%%%%%
\section{Proposed Polar-Domain Channel Estimation}

In this section, we introduce the MRDN, P-MRDN, and P-MSRDN for the CE in XL-MIMO systems.
The MRDN architecture is introduced as the fundamental component of our proposed CE schemes.
Then, we propose the P-MRDN, which can effectively exploit the polar-domain channel sparsity in XL-MIMO.
In addition, we highlight the differences between the MRDN and P-MRDN caused by the  channel sparsity in different domains.
Finally, the proposed P-MSRDN combines the application of the MRDN and ASPP to further improve the CE accuracy.

 %\vspace{-0.3cm}
\subsection{MRDN Architecture}

As shown in Fig. 2 (a), residual dense network (RDN) and convolutional block attention module (CBAM) \cite{[17]} are the building modules of the MRDN \cite{[8]}.

\subsubsection{Input Layer}

By assuming that the real and imaginary parts of the signal $ {\rm {\bf y}} \in \mathbb{C}^{N \times 1}$ are independent,
%After the correlation operation, the received signal $ {\rm {\bf z}} \in \mathbb{C}^{N \times 1}$ is a vector.
we structure them into a matrix $ {\rm {\bf Y}} \in \mathbb{R}^{N \times 2} $.
Thus, the matrix $ {\rm {\bf Y}} $ can be treated as a two-dimensional image and serve as the input of our schemes.

\subsubsection{Basic Structure}

We denote {\it Convolution} and {\it Rectified Linear Units} by ``{\it Conv}" and ``{\it ReLU}", respectively.
``{\it Conv}" and ``{\it ReLU}" layer functions are denoted by $ * $ and max, respectively.
%Assuming that $ * $ denotes ``{\it Conv}" layer function, max denotes ``{\it ReLU}" layer function,
Then, as shown in Fig. 3 (a), the model of the $n$-th residual block is a combination of two cascaded functions:
%$      {\rm {\bf r}}_{-1} = W_{n,r} * {\rm {\bf x}} + b_{n,r}       $ and $  {\rm {\bf r}}_{0} = {\rm max}(0,{\rm {\bf r}}_{-1}) $,
\begin{align}
  {\rm {\bf r}}_{-1} &= W_{n,r} * {\rm {\bf x}} + b_{n,r}, \\
  {\rm {\bf r}}_{0} &= {\rm max}(0,{\rm {\bf r}}_{-1}),
\end{align}
where $ \{ W_{n,r}, b_{n,r}\}$, $  n\in \{ 1,2,\cdots,{ M} \} $, denote the weight and bias matrices, respectively.
As shown in Fig. 3 (a), $M$ is the number of layers of RDN.
%where the weight and bias matrices of the $n$-th residual block parameter are denoted by $ \Xi_{n,r} = \{ W_{n,r}, b_{n,r}\},n\in \{ 1,2,\cdots,{ M} \} $ with $M$ being the number of layers of RDN, as shown in Fig. 2 (a). % and $r$ denoting the residual block.
We denote the input and output of the residual block by $ {\rm {\bf x}} $ and $  {\rm {\bf r}} $, respectively.

\subsubsection{RDN Structure}

As shown in Fig. 2 (a), in the $n$-th residual block, with $ f_n $ denoting the recursion function, the recurrence relation is $ {\mathcal{F}}_1 = f_1( {\rm {\bf x}} ) $ and %$ {\mathcal{F}}_n = f_n( {\mathcal{F}}_{n-1},\cdots, {\mathcal{F}}_{1},  {\rm {\bf x}}   ), \forall n \in \left\{ {2, \cdots ,M} \right\}$.
\begin{equation}
{\mathcal{F}}_n = f_n( {\mathcal{F}}_{n-1},\cdots, {\mathcal{F}}_{1},  {\rm {\bf x}}   ), \forall n \in \left\{ {2, \cdots ,M} \right\}.
\end{equation}

\begin{figure}
	\addtocounter{figure}{0} %%%%%%%%%note1
	\centering
	\subfigure[RDN and CBAM system models.]{
		\begin{minipage}[t]{1\linewidth}%%%%%%%%%note2
        \centering
			\includegraphics[width=0.9\linewidth]{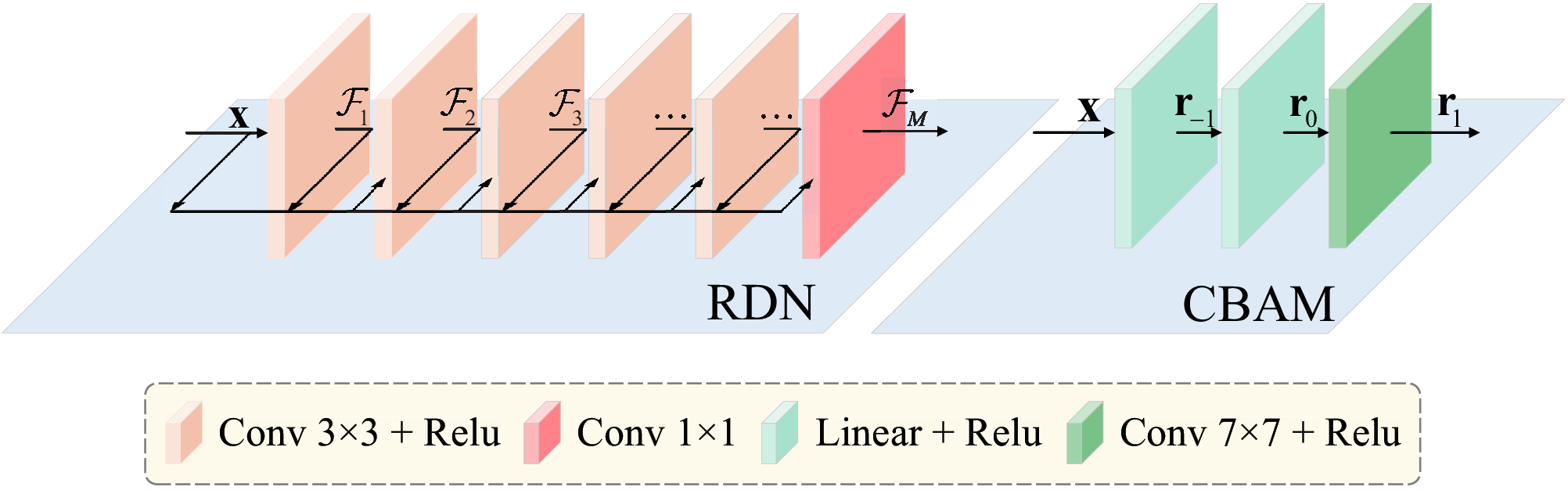}%%%%%%%%%note3
		
		\label{RDN}
		\end{minipage}%
	}%

 %\vspace{-0.1cm}
%%%%%%%%%note4
	\subfigure[ASPP-RDN system model.]{
        \centering
		\begin{minipage}[t]{1\linewidth}
        \centering
				\includegraphics[width=0.8\linewidth]{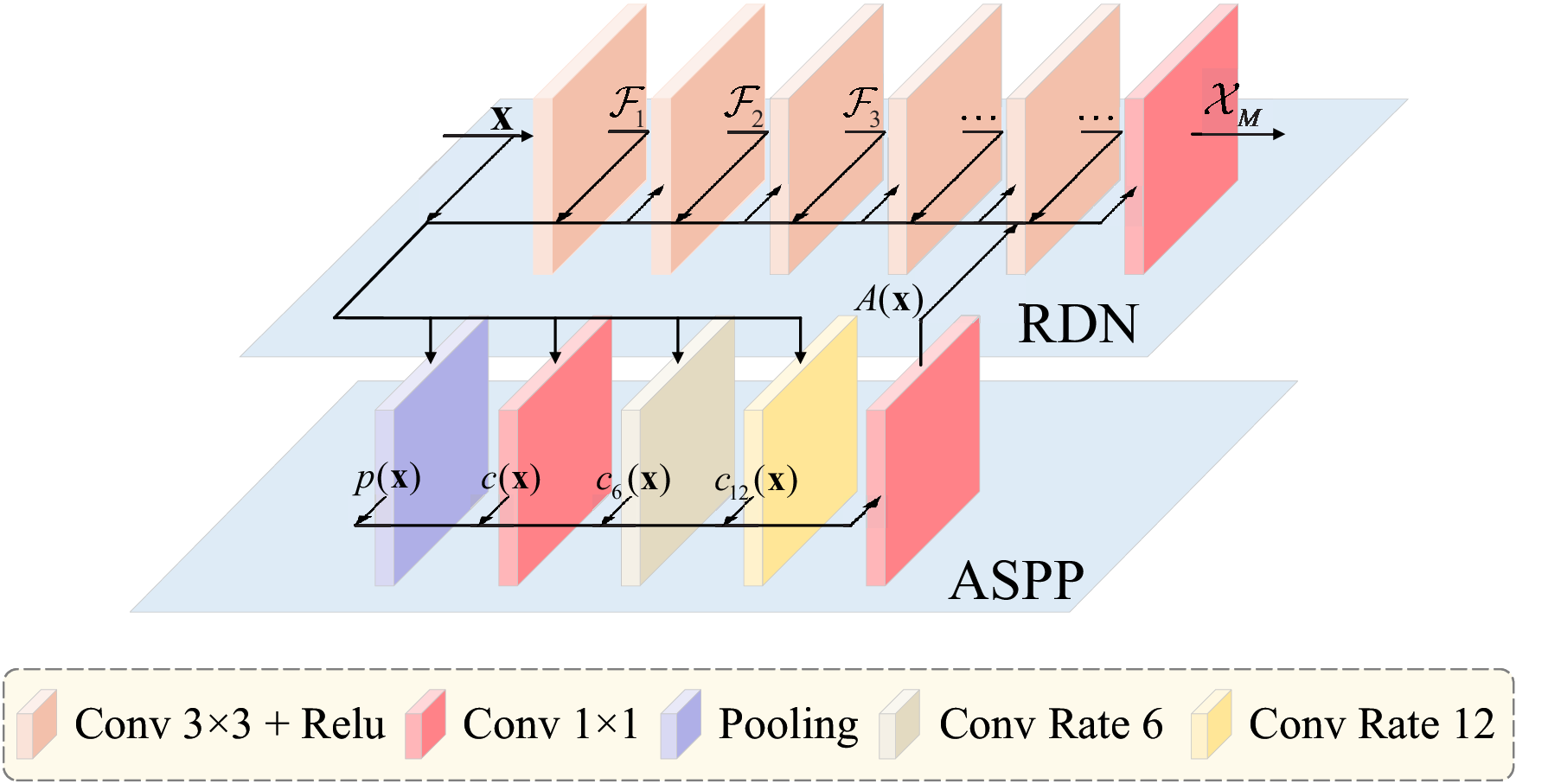}
			
			\label{ASPP-RDN}
		\end{minipage}%
	}%
     % \vspace{-0.4em}
	\caption{Comparison between (a) RDN and CBAM system models and (b) ASPP-RDN system model. The most significant modification of ASPP-RDN is that we take advantages of ASPP as a parallel branch of RDN. The main idea for improvement is to integrate multi-scale features of its input and serve as one of the inputs of the final ``{\it Conv}" layer in RDN.  }
	\centering
  \vspace{-0.3cm}
\end{figure}

\subsubsection{CBAM}

The recurrence relation of the CBAM is
%$ {\rm {\bf r}}_{-1} = W_{-1,c} * {\rm {\bf x}} + b_{-1,c} $, $  {\rm {\bf r}}_{0} = {\rm max}(0,{\rm {\bf r}}_{-1}) $,  $  {\rm {\bf r}}_{1} = W_{1,c} * {\rm {\bf r}}_0 + b_{1,c} $,
\begin{align}
  {\rm {\bf r}}_{-1} &= W_{-1,c} * {\rm {\bf x}} + b_{-1,c}, \\
  {\rm {\bf r}}_{0} &= {\rm max}(0,{\rm {\bf r}}_{-1}),\\
  {\rm {\bf r}}_{1} &= W_{1,c} * {\rm {\bf r}}_0 + b_{1,c},
\end{align}
where $  \{ W_{-1,c}, b_{-1,c}  ,   W_{1,c}, b_{1,c} \} $ are the weight and bias matrices, respectively.
The input and output of the CBAM are denoted by $ {\rm {\bf x}} $ and $  {\rm {\bf r}}_1 $, respectively.

\subsubsection{MRDN Structure}

Assuming that the MRDN consists of $ B $ RDNs and a CBAM, the recurrence relation of the MRDN is %$   {\mathcal{F}}( {\rm {\bf x}} ) = {\mathcal{F}}_{M,B}  \star  {\mathcal{F}}_{M,B-1}  \star  \cdots  {\mathcal{F}}_{M,1}( {\rm {\bf x}} )  $  , $   {\mathcal{M}}( {\rm {\bf x}}  )   =    {\mathcal{F}}  \star C(  {\rm {\bf x}} )    $,
\begin{align}
  {\mathcal{F}}( {\rm {\bf x}} ) &= {\mathcal{F}}_{M,B}  \star  {\mathcal{F}}_{M,B-1}  \star  \cdots  {\mathcal{F}}_{M,1}( {\rm {\bf x}} ),\\
  {\mathcal{M}}( {\rm {\bf x}}  )   &=    {\mathcal{F}}  \star C(  {\rm {\bf x}} ),
\end{align}
where the operate $ \star $ denotes a function composition and $ C(  {\rm {\bf x}} ) $ denotes the recursion function of the CBAM.
%The mapping function of the MRDN can be denoted by $ \mathcal{M}$: $\mathbb{R}^{N \times 2}  \rightarrow   \mathbb{R}^{N \times 2} $.
\subsubsection{Output Layer}
The matrix $     {\rm {\bf \hat H}}  \in  \mathbb{R}^{N \times 2} $  can produce the estimated channel $ {\rm {\bf {\hat h}}}  \in  \mathbb{C}^{N \times 1} $ by reversing the combining in the input layer.
Besides, the loss function is given by %$   \mathcal{L} = {\| {{\rm {\bf {\hat h}}}  - {\rm {\bf h}}} \|^2} $.
\begin{equation}
  \mathcal{L} = {\left\| {{\rm {\bf {\hat h}}}  - {\rm {\bf h}}} \right\|^2} .
\end{equation}

 % \vspace{-0.4cm}
\subsection{P-MRDN-Based Channel Estimation Scheme}

In conventional mMIMO systems, the operating region is far-field, i.e., the distance between the BS and the UE is longer than the Rayleigh distance.
In this case, the channel can be modeled by the planar wave that solely depends on the angular information, resulting in the sparsity in the angular domain.
Moreover, the angular-domain representation $ {\rm  {\bf h} }_{\mathrm{ A}} $ can be converted from the channel $ {\rm  {\bf h} } $ over Fast Fourier Transform (FFT), as discussed above. % with the matrix $ \rm \bf F $

As shown in Fig. 2 (a), the MRDN-based CE scheme aims to fully exploit the channel sparsity in the angular domain by transforming the matrix $ {\rm  {\bf Y} } $ into the angular domain over FFT.
The estimated channel matrix $ {\rm {\bf {\hat H}}} $ is then obtained over Inverse Fast Fourier Transform (IFFT) with the matrix  $    {\rm {\bf \hat H}}_{\mathrm A}  = {\rm {\bf  Y}}_{\mathrm A} - {\mathcal{M}}( {\rm {\bf Y}}_{\mathrm A}  )  $, which can produce the estimated channel $ {\rm {\bf {\hat h}}} $.

%However, in the case of the XL-MIMO, the Rayleigh distance is extended to hectometer-scale distances or beyond, indicating that the operating region is in the near-filed.
%To fully capture the near-field characteristics, XL-MIMO channels should be modeled by the spherical wave, which incorporates both the angle and distance information.
%As a result, the channel sparsity of the XL-MIMO is expressed in the polar domain rather than the angular domain.

To leverage the polar-domain channel sparsity in XL-MIMO systems, the proposed P-MRDN-based channel estimation scheme adopts the polar-domain transform (PT) to transform the matrix $ {\rm {\bf Y} } $ to the polar domain counterpart (i.e., ${\rm {\bf Y} }_{\mathrm{P}}$) through the matrix $ {\rm {\bf D} } $, analogous to the angular domain transformation.
%, which is given in Appendix.
As shown in Fig. 2 (b), the estimated channel $ {\rm {\bf {\hat h}}} $ is constructed by the estimated channel matrix $ {\rm {\bf {\hat H}}} $ which is obtained over the inverse polar-domain transform (IPT) with the matrix $    {\rm {\bf \hat H}}_{\mathrm P}  = {\rm {\bf  Y}}_{\mathrm P} - {\mathcal{M}}( {\rm {\bf Y}}_{\mathrm P}  )  $. %through the matrix $ {\rm  {\bf D} } $.

Once again, the key difference between the MRDN and P-MRDN lies in their approach to exploit the inherent channel sparsity.
The MRDN-based CE scheme transforms the channel to the angular domain, exploiting the angular-domain sparsity in the far-field.
In contrast, the P-MRDN-based CE scheme transforms the channel to the polar domain, leveraging the polar-domain sparsity in the near-field of XL-MIMO.

 %\vspace{-0.4cm}
\subsection{P-MSRDN-Based Channel Estimation Scheme}

To further improve the channel estimation accuracy, we define a parallel part of the ASPP and RDN\footnote{By incorporating the notion of ASPP into the proposed P-MRDN-based CE scheme inspired by \cite{[13]}, advanced deep learning structures can be exploited to improve the performance of the proposed CE schemes. In addition, other similar structures, e.g., Encoder-Decoder and Encoder-Decoder with Atrous Conv \cite{[12]}, can also be employed to improve the performance of the proposed CE schemes. It is crucial to note that introducing more advanced structures is vital in optimizing the complexity, fitting, and generalization capabilities of the proposed schemes. These aspects deserve further investigation in the future.}, named ASPP-RDN, as shown in Fig. 3 (b).
By incorporating the notion of ASPP into the proposed P-MRDN, the new CE scheme can achieve higher NMSE performance as the ASPP can integrate multi-scale features of its input.

\subsubsection{Atrous Spatial Pyramid Pooling Structure}

We denote the single recursion function of the pooling layer,  ``{\it Conv}'' layer and   ``{\it Conv } $ rate = i $'' layer by $ p $, $ c $, and $  c_i $, respectively. Then, the recurrence relation of the ASPP can be given by
%$  A(  {\rm {\bf x}} ) = c (  p( {\rm {\bf x}} ) , c( {\rm {\bf x}} ), c_6( {\rm {\bf x}} ), c_{12}( {\rm {\bf x}} )          ) $,
\begin{equation}
  A(  {\rm {\bf x}} ) = c (  p( {\rm {\bf x}} ) , c( {\rm {\bf x}} ), c_6( {\rm {\bf x}} ), c_{12}( {\rm {\bf x}} )          ),
\end{equation}
where $  {\rm {\bf x}} $ denotes the input of the ASPP, $  A $ is the mapping function for the ASPP.

\subsubsection{ASPP-RDN Structure}

Based on the ASPP and RDN structure, the recurrence relation of the proposed ASPP-RDN is
%$ {\mathcal{X}}  (  {\rm {\bf x}} ) = f_M( A(  {\rm {\bf x}} ) ,{\mathcal{F}}_{M-1},\cdots, {\mathcal{F}}_{1},  {\rm {\bf x}}   ) $,
\begin{equation}
    {\mathcal{X}}  (  {\rm {\bf x}} ) = f_M( A(  {\rm {\bf x}} ) ,{\mathcal{F}}_{M-1},\cdots, {\mathcal{F}}_{1},  {\rm {\bf x}}   ),
\end{equation}
where $ M $ denotes the number of layers of the RDN and $  {\rm {\bf x}} $ denotes the input of the ASPP-RDN.

\subsubsection{P-MSRDN Structure}

The proposed P-MSRDN jointly makes the full use of novel ideas in the ASPP and MRDN, which are illustrated as follows:
\begin{itemize}
  \item MRDN is an extended and versatile architecture of RDN that is cascaded from multiple RDNs. The superiority of the MRDN for CE has been demonstrated due to the similarity between CE and image noise reduction.
  %state-of-the-art for real-world image denoising.
  MRDN is utilized with modifications as the main component of our proposed P-MSRDN.
  \item ASPP has shown excellent performance in reconstructing the texture details while removing the embedded noise. We utilize the ASPP as a parallel branch of the RDN to take advantage of its multi-scale feature integration capabilities, which further enhances the accuracy of CE.
  \item Assuming that there are $ B $ ASPP-RDNs and a CBAM in the proposed P-MSRDN, the recurrence relation of the P-MSRDN is
%  $  \mathcal{Z}( {\rm {\bf x}} ) = \mathcal{X}_{B}  \star  \mathcal{X}_{B-1}  \star  \cdots  \mathcal{X}_{1}( {\rm {\bf x}} ) $, $ \mathcal{Y}( {\rm {\bf x}}  )   =    \mathcal{Z}  \star C(  {\rm {\bf x}} ) $,
  \begin{align}
      \mathcal{Z}( {\rm {\bf x}} ) &= \mathcal{X}_{B}  \star  \mathcal{X}_{B-1}  \star  \cdots  \mathcal{X}_{1}( {\rm {\bf x}} ),\\
      \mathcal{Y}( {\rm {\bf x}}  )   &=    \mathcal{Z}  \star C(  {\rm {\bf x}} ),
  \end{align}
  where $ \mathcal{Y}$: $\mathbb{R}^{N \times 2}  \rightarrow   \mathbb{R}^{N \times 2} $ is the mapping function for the P-MSRDN.
  The final estimated channel  $ {\rm {\bf {\hat h}}}$ can be obtained by reversing the combining in the input layer, as shown in Fig. 2 (b).
  %with  the output $ {\rm {\bf {\hat H}}}  =  \mathcal{Y}( {\rm {\bf Y}}_P  ) $.

\end{itemize}

 %\vspace{-0.3cm}
\subsection{Computational Complexity}

The computational complexity of orthogonal matching pursuit (OMP) and polar-domain orthogonal matching pursuit (P-OMP) are expressed as $ \mathcal{O}(L^3 N^2 ) $ and $ \mathcal{O}(L^3 NQ ) $, respectively \cite{[5]}.
On the other hand, the computational complexity of the running phase in the MRDN is given by \cite{[8]} %$ \mathcal{O} \left(  BM  N^2K^2  E^2    \right) $,
\begin{equation}
  \mathcal{O} \left(  BM  N^2K^2  E^2    \right),
\end{equation}
where $K^2$ is the size of kernels for ``{\it Conv}'' layers and $ E $ denotes the number of features for the MRDN.
In addition, we assume that the number of features for the P-MRDN and P-MSRDN are also $ E $.
The computational complexity of the running phase in the P-MRDN and P-MSRDN are expressed as% $  \mathcal{O} \left(  BM  NQK^2  E^2           \right) $ and $  \mathcal{O} \left( B(M + 4 )   NQK^2 E^2           \right) $, respectively.
\begin{equation}
  \mathcal{O} \left(  BM  NQK^2  E^2           \right),
\end{equation}
and
\begin{equation}
  \mathcal{O} \left( B(M + 4 )   NQK^2 E^2           \right).
\end{equation}

 %\vspace{-0.3cm}
\section{Simulation Result}

%\begin{figure}
%  \centering
%  \includegraphics[width=2.7in]{P-OMP-NMSE}
%  \caption{P-OMP-NMSE.}\label{P-OMP-NMSE}
%\end{figure}

%In this section, simulation results are provided to reveal the superior performance of the proposed channel estimation schemes.
We consider a XL-MIMO system, where $ N = 128 $, $ \lambda = 0.03 $ meters, $ L = 6 $, and $ Q = 256 $.
The complex path gain $ \beta_l $ and the distance $ r_l $ of the $ l $-th path are generated as: $ \beta_l \sim   \mathcal{C}\mathcal{N} (0,1)  $ and $  r_l \sim  \mathcal{U} (5,50)$ meters, respectively.
%In addition, we ignore the large-scale path loss in XL-MIMO.
%In addition, the matrix $ {\rm {\bf D}} $ is generated by the algorithm in \cite{[3]}.
In terms of hardware, we implement the proposed schemes using Intel Core i7-12700, 16 GB RAM, and NVIDIA GeForce GTX 1660 SUPER through PyTorch library.
The learning rate is set as $0.0001$ for the MRDN, P-MRDN, and P-MSRDN.
We adopt $16000$ and $4000$ samples for the training and testing sets of three schemes, respectively.
The number of residual blocks for the RDN and the number of the RDN for our schemes are $6$ and $8$, respectively.
Besides, the NMSE is defined as $
  {\rm  {\bf {NMSE} }} = \mathbb{E} (    {             { \|   {\rm {\bf h}} - {\rm {\bf {\hat h}}} \|^2 }   /          {      \| {\rm {\bf h}} \|^2               }               }).$

\begin{figure}
  \centering
  \setlength{\abovecaptionskip}{-0cm}
  \includegraphics[width=3in]{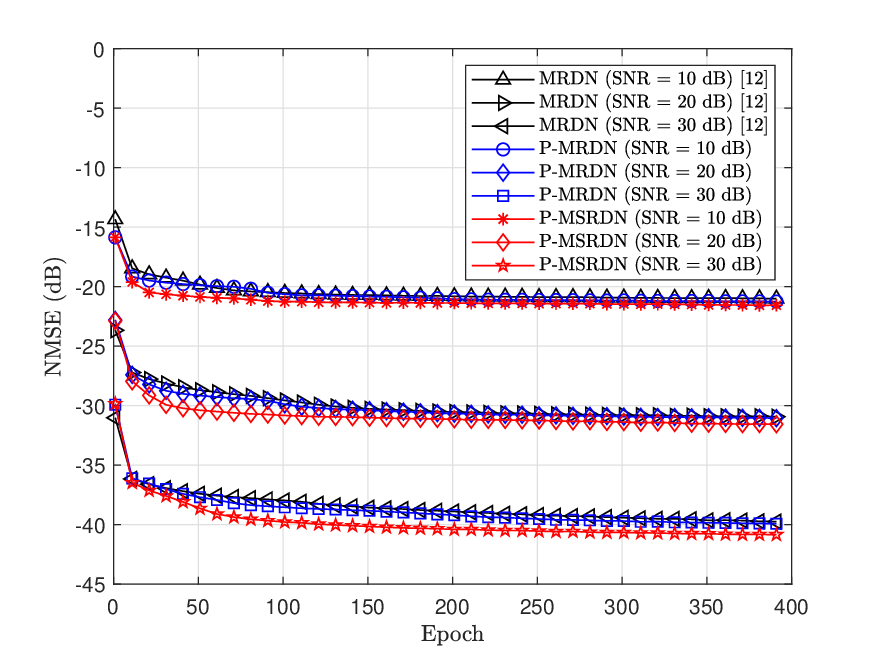}
        %\vspace{-0.4em}
  \caption{Convergence of the three considered channel estimation schemes (i.e., the MRDN, P-MRDN, and P-MSRDN) under different training SNRs.}\label{convergence}
  %\vspace{-0.3cm}
\end{figure}

The convergence performances of the proposed CE schemes are compared in Fig. 4, where the training SNRs are set to $10$, $20$, and $30$ dB, respectively.
The first observation is that the proposed P-MSRDN can achieve the best NMSE performance and the fastest convergence among the considered schemes, irrespective of the training SNRs.
Specifically, after $400$ epochs, the performance gaps between the P-MRDN and MRDN for the training SNRs of $10$, $20$, and $30$ dB are $0.24$, $0.13$, and $0.18$ dB, respectively.
In comparison, the performance gaps between the P-MSRDN and MRDN are $0.55$, $0.63$, and $1.09$ dB, respectively.
The reason for this improvement is that the P-MSRDN can effectively exploit the polar-domain channel sparsity and captures more features compared with the MRDN and P-MRDN.
Furthermore, it is worth noting that all the schemes achieve their convergence within $300$ epochs.

\begin{figure}
  \centering
    \setlength{\abovecaptionskip}{0.cm}
  \includegraphics[width=3in]{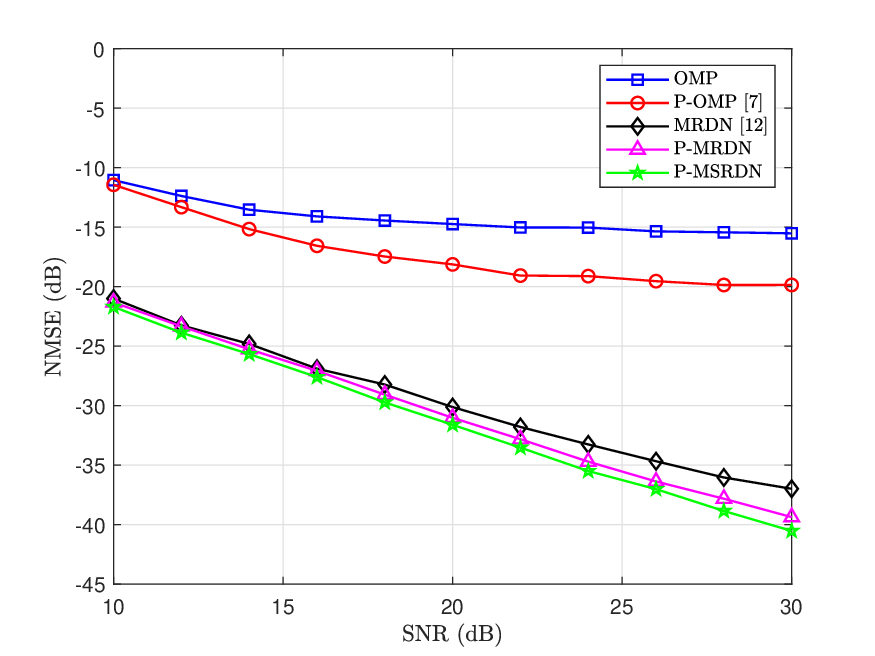}
        %\vspace{-0.4em}
  \caption{NMSE performance comparison of the P-MRDN and P-MSRDN with MRDN and CS schemes.}\label{P-MSRDN-NMSE}
  %\vspace{-0.7cm}
\end{figure}

Fig. 5 compares the NMSE performance of our proposed schemes with the MRDN-based and CS-based schemes.
OMP provides the worst NMSE performance, while P-OMP has a significantly better NMSE performance compared to OMP, with a $3.39$ dB and $4.34$ dB increase when the SNR is $20$ dB and $30$ dB, respectively.
In addition, the proposed P-MRDN outperforms the MRDN by $2.38$ dB when the SNR is $30$ dB.
These reveal the superiority of the polar-domain schemes, which can effectively exploit the rich polar-domain channel sparsity in XL-MIMO for lowering the NMSE in CE.
An interesting finding is that the channel sparsity has minimal influence on the DL-based schemes compared with the CS-based schemes.
The possible reason could be that the MRDN has already learned implicitly a portion of the unstructured near-field information.
It is worth noting that the proposed P-MRDN provides a $12.89$ dB and $19.51$ dB NMSE performance gain over the P-OMP when the SNR is $20$ dB and $30$ dB, respectively.
More importantly, the proposed P-MSRDN can achieve better NMSE performance compared to other schemes, with a gain of $1.49$ dB and $3.54$ dB over the MRDN when the SNR is $20$ dB and $30$ dB, respectively.
We can also find that the NMSE performance gaps between the P-MSRDN and other schemes become larger with the increase of SNR.
Indeed, the exploitation of the ASPP to capture multi-scale features of the polar-domain channel is the main reason for this improvement.
Overall, the proposed P-MSRDN produces the best quality of CE compared to the other schemes.

%\begin{figure}
%  \centering
%    \setlength{\abovecaptionskip}{0.cm}
%  \includegraphics[width=2.8in]{distance}
%        %\vspace{-0.4em}
%  \caption{NMSE performance comparison of the P-MRDN and P-MSRDN with MRDN against the distance between the BS and the UE when the SNR is 20 dB.}\label{distance}
%  \vspace{-0.7cm}
%\end{figure}

%{\textcolor{blue}{Fig. 6 compares the NMSE performance of the P-MRDN, P-MSRDN, and MRDN as a function of the distance between the BS and the UE.
%The SNR is set to $20$ dB.
%At a distance of $10$ meters, we can observe that the proposed P-MSRDN outperforms the MRDN and P-MRDN by $1.94$ dB and $1.17$ dB, respectively.
%Furthermore, the NMSE performance gap between the P-MSRDN and MRDN increases as the distance decreases, demonstrating the superiority of the P-MSRDN, particularly in the case where the distance between the BS and UE will be further reduced.}}

The computational complexity and running time of all the schemes are compared in TABLE \ref{table}.
In particular, we normalize the running time of all the schemes by the one obtained by OMP for comparison.
As discussed above, OMP and P-OMP achieve the lowest computational complexity at the cost of high estimation error.
In addition, all the schemes have the same order of magnitude of running time.
More importantly, the P-MSRDN achieves better NMSE performance compared to the MRDN and P-MRDN, but only requires a similar computational complexity.
This is due to the fact that the P-MSRDN can capture the multi-scale features of the polar-domain channel through the exploitation of the ASPP.

\begin{table}[t]
  \begin{center}

    \caption{Comparison of Computational Complexity and Running time over Different Channel Estimation Schemes.}
          %\vspace{-0.4em}
      \label{table}
    \begin{threeparttable}
    \begin{tabular}{|c|c|c|c|} % <-- Alignments: 1st column left, 2nd middle and 3rd right, with vertical lines in between
      \hline
      \textbf{Scheme} & \textbf{Computational Complexity} & \makecell{ \textbf{Running} \\ \textbf{ Time } \\ \textbf{(ms)}} & \makecell{ \textbf{Ratio of} \\ \textbf{ Running} \\   \textbf{ Time}   }     \\
      \hline
      OMP & $ \mathcal{O}(L^3 N^2 ) $  & 5.47 & $\times$1  \\ %  5.47  \\  3538944( 1 )
      \hline
      P-OMP [7] & $ \mathcal{O}(L^3 NQ ) $&  8.43  & $\times$1.54  \\  %   8.43\\     2
      \hline
      MRDN [12] & $\mathcal{O} \left(  BM  N^2K^2 E^2           \right)$&   4.53 & $\times$0.83  \\ %   8.88\\
      \hline
      P-MRDN & $\mathcal{O} \left(   BM NQK^2 E^2           \right)$& 4.77  &$\times$0.87  \\  %  7.77  \\
      \hline
      P-MSRDN & $\mathcal{O} \left(  B(M + 4 )  NQK^2 E^2           \right)$  & 7.41 &  $\times$1.35 \\ %  9.36\\
      \hline

    \end{tabular}
    %\vspace{0.1cm}
   \begin{tablenotes}[para,flushleft]

         \item In this table, $ N $ is the number of antennas for the BS; $ L $ denotes the number of all path components; $ Q = 2N $ is the number of all the sampled distances for the matrix $ {\rm {\bf D}} $; $ K^2 $ is the size of kernels for ``{\it Conv}'' layers; $B$ denotes the number of RDN for MRDN or ASPP-RDN for P-MSRDN; $ M $ is the number of layers of RDN; $E$ denotes the number of features for ``{\it Conv}'' layers. Besides, the running time ratios of all the schemes are calculated with respect to the one achieved by the OMP and the running time can be further shortened if tailor-mode computing devices are adopted.
        % \item[**] my website is ... %此处加入注释**信息
   \end{tablenotes}
   \end{threeparttable}
  \end{center}
   \vspace{-0.6cm}
\end{table}

%within almost the same computational complexity,
\section{Conclusion}

We proposed the P-MRDN and P-MSRDN-based CE schemes for XL-MIMO systems, building on the conventional MRDN structure.
More specifically, the proposed P-MSRDN can achieve superior generalization capabilities by utilizing several techniques, e.g., exploiting the near-field channel sparsity in the polar domain, deep residual learning, and extracting features at multi-scale resolutions.
By transforming the channel into the polar domain, the proposed P-MRDN and P-MSRDN schemes can effectively exploit the sparsity in the polar domain that outperform the MRDN scheme and the conventional CS schemes.
As for potential future works, the topics could be the CE for hybrid-field scenario of XL-MIMO, where various UEs are in near-field and others are in far-field.

\bibliographystyle{IEEEtran}
\bibliography{IEEEabrv,Ref}

\end{document}